\documentclass[%
nofootinbib,
 amsmath,amssymb,
 aps,pra, twocolumn,
]{revtex4}
\usepackage{relsize}
\usepackage{graphicx}
\usepackage{dcolumn}
\usepackage{xcolor}
\usepackage{bm}
\usepackage{ulem}

\newcommand{\comment}[1]{}
\newcommand{\BEA}{\begin{eqnarray}}
\newcommand{\EEA}{\end{eqnarray}}
\newcommand{\n}{\hat{n}}

\begin{document}


\title{Photon cooling: linear vs nonlinear interactions}


\author{A. Hovhannisyan$^{1}$, V. Stepanyan$^{2}$ and A.E. Allahverdyan$^{3}$}
\affiliation{
$^{1}$Institute of Applied Problems of Physics, 
Yerevan, Armenia\\
$^{2}$Physics Department, Yerevan State University, 
Yerevan, Armenia\\
$^{3}$Alikhanian National Laboratory (Yerevan Physics Institute), 
Yerevan Armenia}

\begin{abstract}
Linear optics imposes a relation that is more general than the second law of thermodynamics: For modes undergoing a linear evolution, the full mean occupation number (i.e. photon number for optical modes) does not decrease, provided that the evolution starts from a (generalized) diagonal state. This relation connects to noise-increasing (or heating), and is akin to the second law and holds for a wide set of initial states. { 
Also, the Bose-entropy of modes increases, though this relation imposes additional limitations on the initial states and on linear evolution.} We show that heating can be reversed via nonlinear interactions between the modes. They can cool|i.e. decrease the full mean occupation number and the related noise|an equilibrium system of modes provided that their frequencies are different. Such an effect cannot exist in energy cooling, where only a part of an equilibrium system is cooled. We describe the cooling set-up via both efficiency and coefficient of performance, and relate the cooling effect to the Manley-Rowe theorem in nonlinear optics. 

\end{abstract}

\maketitle

\section{Introduction}

Cooling is needed for noise-reduction and for capturing quantum degrees of freedom. It has been  studied during the past 100 years in various set-ups \cite{epstein,abragam,walls}. Cooling processes are also fundamental for thermodynamics: they sharpen the understanding of the second law, and are instrumental for the third law \cite{gal-or}. An interesting example of this is the laser cooling of solids via the anti-Stokes effect, which does have both quantum and thermodynamic nature \cite{epstein}. Much attention is currently devoted to cooling processes in quantum thermodynamics 
\cite{mahler,silva,gallego,huber,paz,raeisi,kurizki,uzdin,luis,eco,gonzalez,johal,mosca,huber2,armen,armen2}.
It is known that only a part of a thermally isolated (initially equilibrium) system can be cooled in terms of energy (or temperature), that cooling such systems costs high-graded energy (work), hence the definition of the coefficient of performance (COP), and that cooling is limited by energy spectra and complexity costs.

Here we consider bosonic (for clarity photonic) degrees of freedom (modes), and show that linear transformations (e.g. linear optics) always increase the full photon number of the system. This statement holds for a wide class of initial states. For such states increasing the mean photon number relates to increasing of noise (heating). 
{ The heating is more general than the second law. To confirm this point, we studied the full Bose-entropy of modes. This coarse-grained entropy is conditionally maximal at equilibrium, and can change under a unitary evolution, in contrast to the fine-grained von Neumann entropy. We show that the Bose-entropy can increase, but this relation (a formulation of the second law) demands additional limitations both on the initial states and linear evolution. }

Heating can be reversed by nonlinear interactions. One can cool in this sense an initially {\it equilibrium} system, which consists of two or higher number of modes. This is not possible for energy cooling, where as demanded by the second law, only subsystem's energy can be decreased (cooled). Our cooling set-up is characterized by two efficiency-like parameters: the coefficient of performance (COP) and the efficiency. The former refers to energy costs of cooling, while the latter normalizes the cooling result over the total changes introduced in the system. Nonlinear interactions achieve cooling in near-resonance regimes, where there is an effective conservation law in the number of photons (Manley-Rowe theorem) \cite{weiss,landau}. Thus, this cooling scenario uncovers a thermodynamic role of nonlinear optical processes. We work in terms of photons, but our results hold for other bosons (e.g. phonons).

This paper is organized as follows. The second section shows that the mean boson (photon) number increases in linear evolution if the evolution starts from a certain class of generalized diagonal initial states. This class is sufficiently large and includes usual diagonal states (in the Fock basis), independent states (over the modes) {\it etc}. Section \label{sec2} also relates the increase of the mean number to noise and formulates this as a heating (no-cooling) principle for linear evolution. { 
In Section \ref{entro}, we study the Bose-entropy for modes and explain under what additional restrictions (compared with the mean photon number increase) this entropy grows. This section also addresses the physical meaning of the Bose-entropy.} Section \ref{sec3} describes the optimal cooling set-up for two modes and introduces the basic characteristics of cooling, {\it viz.}, efficiency and the coefficient of performance (or COP). This section emphasizes the key feature of this cooling setup, namely: the global cooling of an equilibrium system in terms of the mean photon number (and noise) is possible. Section \ref{sec4} demonstrates that cooling is possible also via a feasible nonlinear two-mode interaction, works out a simple example of such interactions, and establishes relations with the Manley-Rowe theorem, a known result in nonlinear physics. The last section summarizes our results. 

\section{No cooling for linear interactions} 
\label{sec2}

\subsection{A single bosonic mode}

Linear processes, which are described with Hamiltonians quadratic in creation/annihilation operators, describe the lion's share of boson dynamics \cite{caves,garrison}. Consider the simplest example of such processes: a single mode that underwent a linear evolution governed by a quadratic Hamiltonian. In the Heisenberg picture, the general form of this evolution connects initial $a=a(0)$ and final $b=a(t)$ annihilation operator of the mode:
 \begin{equation}
 \label{1}
     b = S a + R a^\dagger + f,
 \end{equation}
 where $S$, $R$ and $f$ are complex c-numbers that characterize the evolution. The initial state (density matrix) $\rho$ of the mode satisfy:
\BEA
\label{22}
\langle a\rangle\equiv{\rm tr}(a\rho)=\langle a^\dagger\rangle=0. 
\EEA
The commutation relation $[b,b^\dagger]=[a,a^\dagger]=1$ impose $|S|^2 - |R|^2 =1$ in (\ref{1}). Then we get from (\ref{1}, \ref{22})
\begin{equation} 
\label{eq:deltan}
\langle b^\dagger b \rangle - \langle a^\dagger a \rangle = 2 |R|^2 \langle a^\dagger a \rangle + |R|^2 + |f(t)|^2 \geq 0,
\end{equation}
i.e. the mean photon number difference defined in the LHS of (\ref{eq:deltan}) can only increase. In particular, this conclusion holds for linear amplifiers \cite{caves}. 
According to (\ref{1}) also the dispersion of the photon number increases: 
\BEA
\langle (b^\dagger b)^2\rangle-\langle b^\dagger b\rangle^2\geq \langle (a^\dagger a)^2\rangle-\langle a^\dagger a\rangle^2. 
\EEA
We emphasize that the analogue of (\ref{1}, \ref{eq:deltan}) for a fermion mode does not generally hold. One heuristic reason for this is that only for the bosonic mode the mean (photon) number can be arbitrary large.

\subsection{Many modes, relations with noise and heating}
\subsubsection{Linear Heisenberg evolution}

Importantly, (\ref{eq:deltan}) extends to the completely general $N$-mode situation, where instead of (\ref{1}) we write for initial $a_i=a_i(0)$ and final $b_i=a_i(t)$ Heisenberg operators
 \begin{equation}\label{eq:multim}
     b_i = {\sum}_{j=1}^N \left( S_{ij} a_j + R_{ij} a^\dagger_j\right)+f_i, \quad i = 1, \dots, N,
 \end{equation}
 where $S_{ij}$, $R_{ij}$ and $f_i$ are c-numbers; cf.~(\ref{1})). Write (\ref{eq:multim}) in block-matrix form:
 \begin{gather}\label{eq:matrixform}
 \begin{pmatrix} b \\ b^\dagger \end{pmatrix} = E \begin{pmatrix} a \\ a^\dagger \end{pmatrix}
 + \begin{pmatrix} f \\ f^* \end{pmatrix}, \quad E = \begin{pmatrix} S & R \\ R^* & S^* \end{pmatrix},
\end{gather}
where $a = (a_1, \dots a_N)^T$, $a^\dagger$, $b$, $b^\dagger$, $f$, and $f^*$ are $N$-columns, and where $^T$ and $^*$ denote (resp.) transposition and complex conjugation; below $^{\dagger}=^{*T}$ will denote hermitean conjugation.

Now commutation relations $[b_i, b_j^\dagger]= [a_i, a_j^\dagger] = \delta_{ij}$, where $\delta_{ij}$ is the Kronecker's delta, and $[b_i, b_k]=[a_i, a_k] = 0$ lead from (\ref{eq:matrixform}) to (resp.):
\BEA
\label{2}
 SS^\dagger - R R^\dagger = I, \quad S R^T = R S^T,
\EEA
where $I$ is the $N\times N$ unit matrix. Eqs.~(\ref{2}) imply 
 \begin{gather}\label{eq:matrixinverse}
 E^{-1} = \begin{pmatrix} S^\dagger & -R^T \\ -R^\dagger & S^T \end{pmatrix}.
\end{gather}
The reasoning that led to (\ref{2}) is now applied to (\ref{eq:matrixinverse}), since the same commutation relations hold. Then we get in addition to (\ref{2}) the following new relations:
\BEA
\label{3}
S^\dagger S - R^T R^* = I, \quad S^\dagger R = R^T S^*.
\EEA

\subsubsection{The initial state}

Now assume that the initial state $\rho$ of $N$ modes holds the following two conditions:
\BEA
\label{4}
\langle a_j\rangle\equiv {\rm tr}(\rho a_j)=0,\\
\langle a_i a_j\rangle\equiv {\rm tr}(\rho a_ja_i)
=0,
\label{44}
\EEA
where $i,j = 1, \dots, N$. Two interesting examples of (\ref{4}, \ref{44}) are as follows. First, (\ref{4}) can refer to initially independent modes in states with $\langle a_i\rangle=0$. Then (\ref{44}) holds automatically due to the independence: 
\BEA
\langle a_i a_j\rangle=\langle a_i\rangle\langle a_j\rangle=0. 
\label{indo}
\EEA
Second, we can consider diagonal states $\rho_{\rm diag}$ that read in the Fock basis 
\BEA
\label{diag}
&&\rho_{\rm diag}=\sum_{\nu_1,...,\nu_N=0}^\infty
r_{\nu_1,...,\nu_N}|{\nu_1,...,\nu_N}\rangle\langle
{\nu_1,...,\nu_N}|,\\
&&a_i^\dagger a_i |{\mu_1,...,\mu_N}\rangle=
\mu_i|{\mu_1,...,\mu_N}\rangle,
\label{fo}
\EEA
where (\ref{fo}) defines the Fock basis, and 
where (\ref{diag}) ensures conditions (\ref{4}, \ref{44}). It should be clear that neither independence nor diagonality is necessary for the validity of
(\ref{4}, \ref{44}); e.g. non-diagonal state holding (\ref{4}, \ref{44}) can be easily constructed starting from (\ref{diag}). To be concise, we will refer to the states $\rho$ satisfying (\ref{4}, \ref{44}) as generalized diagonal states.

\subsubsection{Increase of the mean photon number}

Using (\ref{4}, \ref{44}) together with the first equation in (\ref{3}) we find that the change of the total occupation number is non-negative:
\begin{gather}
{\sum}_{i=1}^N \left(
\langle b^\dagger_{i}b_{i} \rangle-\langle a^\dagger_{i}a_{i} \rangle\right)
={\sum}_{i=1}^N |f_i|^2+{\sum}_{i,j=1}^N |R_{ij}|^2
\nonumber\\
\label{garu}
+2{\sum}_{i=1}^N{\rm tr}\left(Y_i\rho Y_i^\dagger\right)
\geq 0,
\end{gather}
where we defined
\begin{gather}
Y_i\equiv{\sum}_{k=1}^N R^*_{ik}a_k.
\end{gather}

\comment{
If we additionally impose condition 
$\langle a^\dagger_i a_j\rangle=
\delta_{ij}\langle a^\dagger_i a_i\rangle$ and use it together with (\ref{4}, \ref{44}), then a simpler expression is found instead of (\ref{garu}):
\begin{gather}
{\sum}_{i=1}^N \left(
\langle b^\dagger_{i}b_{i} \rangle-\langle a^\dagger_{i}a_{i} \rangle\right)
={\sum}_{i=1}^N |f_i|^2
\nonumber\\
+{\sum}_{i,j=1}^N (2\langle a^\dagger_{i}a_{i} \rangle+1) |R_{ij}|^2
\geq 0.
\label{garuu}
\end{gather}
}

When deducing (\ref{garu}), condition (\ref{4}) was needed for nullifying terms $\propto f_i\langle a_k\rangle$ in (\ref{garu}), while (\ref{44}) was needed for nullifying terms $\propto (R^\dagger S)_{kl}\langle a_ka_l\rangle$.

Eq.~(\ref{eq:multim}) can describe absorption (attenuation) of photons from a few selection target modes, at the expense of their overall increase.  For the particular case of Gaussian initial states, (\ref{garu}) follows from the result of Ref.~\cite{karen} on the maximal work. Thus,  according to (\ref{garu}) the full mean photon number can only increase under linear evolution. 

{ Where these additional photons come from? Answering this question is contingent on realization of the linear transformation. For example, the genesis of additional photons is relatively clear when the increase of the mean number of photons is accompanied by an increase in the overall mean energy; cf.~(\ref{eq:deltan}). This energy increase comes from external sources that realize the linear dynamics. In particular, this is the case when the $N$ modes start their evolution from the overall vacuum state, because then the mean energy can only increase. More generally, the relation between the mean energy increase and the mean photon number increase in a linear dynamics is absent: the latter is more general than the former; see (\ref{oso}) for clarification. In such cases the genesis of additional photons should be prescribed to the general fact that the mean photon number is not conserved within linear dynamics.}

\subsubsection{Noise increase and heating}

We emphasize that (\ref{garu}) can be interpreted as uncertainty increase. To this end, let us note, for a mode with annihilation operator $a$, that $\langle a^\dagger a \rangle$ characterizes the dispersion $\langle\Delta a^2\rangle$ of $a$ \cite{caves}:
\BEA
\label{de}
\langle\Delta a^2\rangle\equiv \frac{1}{2}\langle a a^\dagger+a^\dagger a \rangle-|\langle a \rangle|^2
=\langle a^\dagger a \rangle+\frac{1}{2}-|\langle a \rangle|^2~~~~~~\\
=\langle x^2\rangle-\langle x\rangle^2+\langle y^2\rangle-\langle y\rangle^2,\quad a=x+i y,
~~~~~~~~~~
\label{ded}
\EEA
where $x=(a+a^\dagger)/2$ and $y$ are Hermitian operators. Eq.~(\ref{de}) is the definition of dispersion for non-hermitian $a$, while (\ref{ded}) shows how it can be measured via its Hermitian components $x$ and $y$. 
Note from (\ref{de}) that for $\langle a \rangle=0$, the dispersion $\langle\Delta a^2\rangle$ reduces to 
the mean photon number $\langle a^\dagger a \rangle$ \footnote{This quantity also controls the shot noise in photodetection \cite{garrison}.}. 

For considered initial states (\ref{4}), we have $\langle a_i\rangle=\langle b_i\rangle=0$, and then (\ref{garu}, \ref{ded}) imply that also the sum of uncertainties (\ref{de}) increases together with the photon number:
\begin{gather}
\label{ord}
    {\sum}_{i=1}^N \left(
\langle \Delta b^2_{i} \rangle-\langle \Delta a^2_{i} \rangle\right)=
{\sum}_{i=1}^N \left(
\langle b^\dagger_{i}b_{i} \rangle-\langle a^\dagger_{i}a_{i} \rangle\right)\geq 0,
\end{gather}
i.e. as the mean photon number rises, so does the total dispersion. Eq.~(\ref{ord}) holds due to initial conditions (\ref{4}, \ref{44}) and will be interpreted as heating. Likewise, the decrease of both quantities in (\ref{ord})|that is possible due to nonlinear interactions|will mean cooling; see below. 

{ We close this part by stressing that the relation between ${\sum}_{i=1}^N \left( \langle \Delta b^2_{i} \rangle-\langle \Delta a^2_{i} \rangle\right)$ and ${\sum}_{i=1}^N \left( \langle b^\dagger_{i}b_{i} \rangle-\langle a^\dagger_{i}a_{i} \rangle\right)$ is not automatic. For example, linear dynamics under a particular condition $f_i=0$ in (\ref{eq:multim}) will hold (\ref{garu}) under condition (\ref{44}) only, i.e. (\ref{4}) is now not needed. And then if $\langle a_i\rangle\not =0$, then generically also $\langle b_i\rangle\not =0$ and ${\sum}_{i=1}^N \left( \langle \Delta b^2_{i} \rangle-\langle \Delta a^2_{i} \rangle\right)\geq 0$ does not hold, though ${\sum}_{i=1}^N \left( \langle b^\dagger_{i}b_{i} \rangle-\langle a^\dagger_{i}a_{i} \rangle\right)\geq 0$ still holds due to $f_i=0$.}

{ 
\section{Entropic formulation of the second law for bosons}
\label{entro}

\subsection{When Bose-entropy increases for a linear dynamics?}

\subsubsection{Definition of Bose-entropy }

Eq.~(\ref{garu}) shows that for initial conditions (\ref{4}, \ref{44}) the total mean number of photons can only increase. In the context of this unidirectional change it is natural to ask whether one can find a suitable entropy function that also increases under linear dynamics. As we show below, the answer to this question is positive provided the initial states and the type of the linear dynamics are restricted.

First of all, we need to define the entropy function: as always with the unitary dynamics the von Neumann entropy $-{\rm tr}(\rho\ln\rho)$ (with $\rho$ being the density matrix) is not suitable for defining the second law, since it is conserved. We need a more coarse-grained (i.e. less microscopic) definition of entropy. A good choice is the time-dependent Bose entropy 
\begin{gather}
\label{uu1}
S(t)={\sum}_{k=1}^N s(n_k(t)), \quad n_k(t)\equiv\langle a^\dagger_k(t)a_k(t)\rangle, \\ 
s(n_k)\equiv (1+n_k)\ln [1+n_k]-n_k\ln [n_k].
\label{uu2}
\end{gather}
Eq.~(\ref{uu1}) is deduced for an ideal Bose gas from the microcanonic distribution \cite{landau}. If $s(n_k)$ from (\ref{uu2}) is maximized for a fixed mean energy $\hbar\omega_kn_k$ of the mode $k$ with frequency $k$, one obtains the thermal expression for the mean occupation (photon) number. Indeed, making the Lagrange function $s(n_k)-\beta\hbar\omega_kn_k$, where $\beta$ is the Lagrange multiplier (inverse temperature) one obtains $n_k=(e^{\beta\hbar\omega_k}-1)^{-1}$. Eq.~(\ref{uu1}) also increases in time within kinetic equations for weakly interacting bosons; see \cite{bose-t} for a recent discussion. 

\subsubsection{Increase of Bose-entropy }

To study the behavior of $S$ in time for our situation, we need to add an additional initial condition in (\ref{4}, \ref{44})
\BEA
\label{444}
\langle a^\dagger_i a_j\rangle= \delta_{ij}\langle a^\dagger_i a_i\rangle,
\EEA
where (\ref{444}) holds for examples (\ref{indo}, \ref{diag}). Without (\ref{444}), i.e. staying with (\ref{4}, \ref{44}) only, we cannot express $n_k(t)$ via $n_k(0)$. Together with (\ref{444}) this task is possible from (\ref{eq:multim}):
\begin{gather}
n_i(t)=
\langle b^\dagger_{i}b_{i} \rangle
={\sum}_{k=1}^N (|S_{ik}|^2+|R_{ik}|^2) n_k(0)
\\
+{\sum}_{k=1}^N |R_{ik}|^2
+{\sum}_{i=1}^N |f_i|^2,
\label{garuu}
\end{gather}
where (\ref{2}) and (\ref{3}) imply
\begin{gather}
\label{er1}
 {\sum}_{k=1}^N(|S_{ik}|^2+ |R_{ik}|^2)=1+2{\sum}_{k=1}^N |R_{ik}|^2\geq 1,\\
 {\sum}_{i=1}^N(|S_{ik}|^2+ |R_{ik}|^2)=1+2{\sum}_{i=1}^N |R_{ik}|^2\geq 1. 
\label{er2}
\end{gather}

Let us assume that (consistently with (\ref{er1}, \ref{er2})) there exists a double stochastic matrix $\Theta_{ik}$, i.e. a matrix holding
\BEA
\label{ds}
\Theta_{ik}\geq 0, \quad {\sum}_{i=1}^N\Theta_{ik}= 1,\quad {\sum}_{k=1}^N\Theta_{ik}=1,
\EEA
such that \footnote{For the validity of (\ref{gori}) we in fact need instead of (\ref{ds}) a seemingly weaker condition, where ${\sum}_{i=1}^N\Theta_{ik}= 1$ in (\ref{gori}) is replaced by ${\sum}_{i=1}^N\Theta_{ik}\geq 1$. However, this condition together with ${\sum}_{k=1}^N\Theta_{ik}=1$ and $\Theta_{ik}\geq 0$ leads to ${\sum}_{i=1}^N\Theta_{ik}= 1$. }
\BEA
\label{dss}
|S_{ik}|^2+ |R_{ik}|^2\geq \Theta_{ik}.
\EEA
Matrices $|S_{ik}|^2+ |R_{ik}|^2$ that satisfy (\ref{dss}) are called double-superstochastic \cite{marshall,olkin}. 
Once (\ref{ds}, \ref{dss}) are assumed, the derivation of the second law in the Bose-entropic formulation becomes straightforward from noting that $s(n_k)$ from (\ref{uu2}) is a positive, increasing and concave function:
\begin{gather}
S(t)={\sum}_{i=1}^N s(n_i(t))\geq {\sum}_{i=1}^N s\left [ {\sum}_{k=1}^N \Theta_{ik} n_k(0)\right]\nonumber\\
\geq {\sum}_{i,k=1}^N \Theta_{ik} s[n_k(0)]={\sum}_{k=1}^N s[n_k(0)]=S[0].
\label{gori}
\end{gather}
Thus initial conditions (\ref{4}, \ref{44}, \ref{444}) and dynamic restriction (\ref{dss}) are sufficient for the second law (\ref{gori}). 

\subsubsection{Validity of inequality (\ref{dss})}

Note that (\ref{dss}) trivially holds for $|R_{ik}|^2=0$. 
We emphasize that (\ref{dss}) implies (\ref{er1}, \ref{er2}), but the converse does not hold. To avoid confusions note that ${\sum}_{k=1}^N(|S_{ik}|^2+ |R_{ik}|^2)\leq 1$ and ${\sum}_{i=1}^N(|S_{ik}|^2+ |R_{ik}|^2)\leq 1$ do imply $|S_{ik}|^2+ |R_{ik}|^2\leq \Theta_{ik}$ for some double-stochastic matrix $\Theta_{ik}$ \cite{marshall,olkin}. 

Inequality (\ref{dss}) holds for $N=2$; see Appendix \ref{ap0} which also discusses the simplest counter-example of (\ref{dss}) for $N=3$. A constructive necessary and sufficient condition for the validity of (\ref{dss}) was found in \cite{cruse}:
\BEA
{\sum}_{i\in {\cal I},\,k\in {\cal J}}(|S_{ik}|^2+ |R_{ik}|^2)\geq |{\cal I}|+|{\cal J}|-N,
\label{lolo}
\EEA
where (\ref{lolo}) should hold for {\it all} subsets ${\cal I}$ and ${\cal J}$ of $\{1,...,N\}$, and where
$|{\cal I}|$ and $|{\cal J}|$ are the number of elements in (resp.) ${\cal I}$ and ${\cal J}$. Conditions (\ref{lolo}) are straightforward to check at least for not very large $N$. The physical meaning of (\ref{lolo}) is that sufficiently small values of $(|S_{ik}|^2+ |R_{ik}|^2)$ are to be excluded. 

More general (but less constructive) sufficient dynamical conditions for (\ref{gori}) can be stated as well. For example, whenever (\ref{dss}) does not hold, but still 
\BEA
s\left [ {\sum}_{k=1}^N (|S_{ik}|^2+|R_{ik}|^2) n_k(0)\right.\nonumber\\ \left.
+{\sum}_{k=1}^N |R_{ik}|^2 +{\sum}_{i=1}^N |f_i|^2
\right]\nonumber\\
\geq {\sum}_{k=1}^N (|S_{ik}|^2-|R_{ik}|^2) s[n_k(0)],
\label{nor}
\EEA
holds for all $i$, we sum both parts of (\ref{nor}) over $i$, employ (\ref{er2}) and find
$S(t)\geq S(0)$. 

\subsection{Similarities and differences with the standard formulation of the second law}
\label{sl}

We found two unidirectional relations inherent in linear dynamics for bosons: inequality (\ref{garu}) states on a  increase of the mean photon number, while (\ref{gori}) is about the increase of the Bose-entropy. It is useful to compare these relations with the standard (Thomson's) formulation of the second law \cite{lindblad,thomson}: a unitary dynamics does not decrease the mean energy of a quantum system that started its evolution from a Gibbsian equilibrium (or at least passive) state. The unitary dynamics is realized via time-dependent, cyclically changing Hamiltonian; the cyclic condition is needed for ensuring that the initial and final Hamiltonians are equal \cite{lindblad,thomson}.

Similarities:

-- Eqs.~(\ref{garu}, \ref{gori}) and Thomson's formulation refer to unidirectional changes inherent in a unitary evolution. All of them hold for specific initial states. 

-- For the single-mode situation (\ref{garu}) [i.e. (\ref{eq:deltan})] refers to the basically same quantity as the Thomson's formulation, since the mean photon number is just proportional to the mean energy. 

-- Eq.~(\ref{garu}) relates to noise increase; cf.~(\ref{de}, \ref{ded}). The same holds for the entropic formulation (\ref{gori}) that refers to the Bose-entropy (\ref{uu1}). Thomson's formulation has a similar bridge, since it also tells about the broadening of the energy distribution in the final state as compared to the initial state. This broadening is quantified by the entropy of the energy probability distribution \cite{lindblad, thomson}.

Differences:

$\bullet$ The second law holds for any unitary evolution, while (\ref{garu}) is restricted to a linear evolution of boson modes. Inequality (\ref{gori}) assumes even more restriction; see (\ref{dss}) and (\ref{444}).

$\bullet$ The direct relation between the energy and photon number is broken for the multimode situation, i.e. the analogue of (\ref{garu}) for energy does not hold: the mean energy change
\BEA
\label{oso}
{\sum}_{i=1}^N\omega_i \left(\langle b^\dagger_{i}b_{i} \rangle-\langle a^\dagger_{i}a_{i} \rangle\right), 
\EEA
need not have a definite sign for initial conditions (\ref{4}, \ref{44}). For (\ref{oso}), the derivation that led to (\ref{garu}) breaks down at the point when after the summation over index $i$, one needs to employ the first equation in (\ref{3}). The same holds for (\ref{nor}): it does not apply to the mean energy. In other words, (\ref{nor}) states that the Bose-entropy must increase without simultaneously increasing the mean energy (or at least keeping it constant). 

$\bullet$ Applicability domain: the second law demands equilibrium (e.g. Gibbsian), or at least passive initial state \cite{lindblad,thomson}, while (\ref{4},\ref{44}) and (\ref{22}) allow  initial states that need not be equilibrium or passive; cf.~(\ref{diag}). Recall that a passive state has a density matrix $\rho$ that a non-increasing function of the Hamiltonian $H$ \cite{lindblad,thomson}. For a (Gibbsian) equilibrium state this function is specific: $\rho=e^{-\beta H}/{\rm tr}e^{-\beta H}$ with $\beta>0$ being the inverse temperature \cite{lindblad,thomson}. Thus, (\ref{garu}) is more general than the second law in the context of initial states, but at the same time it is less general in the context of dynamics, as it is restricted to linear evolution. Inequality (\ref{gori}) assumes more restriction on the initial state; see (\ref{444}).
} 

\section{Cooling two equilibrium modes} 
\label{sec3}

\subsection{Set-up}

Once (\ref{ord}) is understood to define heating for linear dynamics with initial conditions (\ref{4}, \ref{44}), it is natural to ask whether non-linear processes can cool, i.e. decrease the initial number of photons. To facilitate the thermodynamic meaning of this question, we shall consider two initially Gibbsian equilibrium bosonic modes at the same temperature $T$. Now a single equilibrium mode cannot be cooled by any unitary (generally nonlinear) operation, since the mean occupation number is proportional to the energy, and the energy decrease for such a situation is prohibited by the second law. However, two initially equilibrium modes at different frequencies can be cooled, in terms of the mean full occupation number, via specific non-linear interactions. Hence, we shall first determine the optimal cooling, and then turn to non-optimal but feasible scenario from the viewpoint of experimentally realizable nonlinear interactions.

Consider the initial state of two modes with frequencies $\omega_1$ and $\omega_2$ at temperature $T$:
\begin{gather}
\label{in}
    \rho = \xi\, e^{-\beta \sum_{i=1}^2\omega_i\hat{n}_i},\quad 
    \xi=(1-e^{-\beta\omega_1})
    (1-e^{-\beta\omega_2}), \\
    \hat{n}_i\equiv a^\dagger_i a_i,\quad
    i=1,2,\quad 
    \hat{n}\equiv {\sum}_{i=1}^2 a^\dagger_i a_i,
\label{def}
\end{gather}
where $\hbar=1$, $\beta = 1 / (k_{\rm B} T) $ and ${\n}_i$ are the occupation number operator for each mode. The two-mode system undergoes a unitary process that aims at cooling:
\BEA
\rho(t)=U\rho\,U^{\dagger},\qquad U U^{\dagger}=1.
\label{star}
\EEA

\subsection{COP and efficiency} 

Besides targeting the mean occupation number, we characterize the cooling via two efficiency-like quantities. Since $\rho$ in (\ref{in}) is an equilibrium state, the final average energy found from (\ref{star}) is larger than the initial one, which is the second law:
\begin{gather}
\label{gulg}
{\sum}_{i=1}^2\omega_i \Delta n_i\geq 0,\quad
\Delta n_{i} \equiv \text{tr}(\rho [U^\dagger \hat{n}_i U -\hat{n}_i ]).
\end{gather}
Eq.~(\ref{gulg}) defines the energy cost of cooling and it motivates the usual definition of coefficient of performance (COP) \cite{armen2}, where the achieved cooling $-{\sum}_{i=1}^2\Delta n_i>0$ is divided over the energy cost ${\sum}_{i=1}^2\omega_i \Delta n_i$. 

Let us define the frequency ratio as 
\BEA
\label{ratio}
\alpha \equiv \frac{\omega_2 }{ \omega_1}<1.
\EEA
We use the dimensionless COP (coefficient of performance) conventionally defined as:
\begin{equation}\label{cop2}
        K = -\frac{\Delta n_1 + \Delta n_2}{\Delta n_1 + \alpha\Delta n_2},  
\end{equation}
where a larger $K$ means e.g. a better cooling with a smaller energy cost. In (\ref{cop2}) we took $\alpha<1$ without loss of generality. Hence, the fact of cooling $-{\sum}_{i=1}^2\Delta n_i>0$ implies via (\ref{gulg}) and $\alpha<1$
\BEA
0\leq \alpha(-\Delta n_2)\leq \Delta n_1
\leq (-\Delta n_2).
\label{raut}
\EEA
Now (\ref{raut}) motivates us to define $\Delta n_1-\Delta n_2=|\Delta n_1|+|\Delta n_2|$ as the total number of occupation changes introduced in the system. This is consistent with thinking about the cooling as photon conversion: some amount of low energy photons ($\Delta n_2<0$) transform into a smaller amount of higher energy photons ($\Delta n_1>0$). The sum of low energy photons given and high energy photons received will be the total number of occupation changes.  Only a fraction $\eta$ of those lead to cooling:
    \begin{equation}\label{eff}
        \eta = -\frac{\Delta n_1 + \Delta n_2}{\Delta n_1 - \Delta n_2}.
    \end{equation}
We call $\eta$ the efficiency of cooling. It is similar to other quantum efficiencies employed in optics \cite{garrison,walls}. Using (\ref{gulg}, \ref{raut}) we get a bound where temperatures are replaced by frequencies: \begin{equation}
\label{carnot}
    \eta \leq \frac{\Delta n_1 + \Delta n_2}{\Delta n_2} \leq 1- \frac{\text{min}[\omega_1,\omega_2]}{\text{max}[\omega_1,\omega_2]},
\end{equation}
i.e. cooling is impossible for $\omega_1=\omega_2$. { Note that (\ref{carnot}) is more similar to the Otto efficiency than to the Carnot efficiency of heat-engines \cite{johal2}.}

\subsection{ Optimal cooling} 

Given (\ref{star},\ref{def}), we look for the unitary which minimizes the mean of $\n$ in the final state:
\BEA
\label{sham}
    U_{\rm opt} = {\rm argmin}_{U}\,[\,
    {\rm tr}(U\rho U^\dagger \hat{n})],
\EEA
Noting the eigenresolutions [cf.~(\ref{in}, \ref{def})]
\begin{equation}
        \rho = {\sum}_{k=0}^\infty r_k |r_k\rangle \langle r_k|,\quad
        \n= {\sum}_{l=0}^\infty \nu_l |\nu_l\rangle \langle \nu_l|,
        \label{ko}
\end{equation}
we get from (\ref{star}, \ref{ko}, \ref{sham}) 
\BEA
{\rm tr}(U\rho U^\dagger \hat{n}) =
{\sum}_{k,l=0}^\infty r_k \nu_l z_{kl}, ~ z_{kl}=
|\langle \nu_l| U |r_k\rangle|^2,~~
\label{min_uni}
\EEA
where 
\BEA
{\sum}_k z_{kl} = {\sum}_l z_{kl} = 1, 
\EEA
i.e.  $z_{km}$ is a doubly stochastic matrix; cf.~(\ref{ds}). Such matrices form a compact convex set with vertices being permutation matrices \cite{olkin}. As (\ref{min_uni}) is linear over $z_{km}$, it reaches the minimum value on the vertices, i.e. on permutation matrices $z_{kl}$. This implies from (\ref{min_uni}) that $U_{\rm opt}$ can be chosen as a permutation matrix.

Thus $U_{\rm opt}$ is a permutation matrix, and its form is seen from (\ref{min_uni}, \ref{ko}):
\BEA
\label{coltrane}
\text{min}_{U}[\,
    {\rm tr}(U\rho U^\dagger {\hat{n}})\,]={\sum}_{k=0}^\infty \nu_k^{\uparrow}r_k^{\downarrow}, \\
    \nu_1^{\uparrow}\leq \nu_2^{\uparrow}\leq \nu_3^{\uparrow} ..., \qquad 
    r_1^{\downarrow}\geq r_2^{\downarrow}\geq
    r_3^{\downarrow} ...,
    \label{ohta}
\EEA
where in (\ref{ohta}) [cf.~(\ref{ko})] the ordered
(anti-ordered) eigenvalues of $\n$ ($\rho$) refer to the final state in (\ref{star}). We visualize the orderings of eigenvalues in the initial state (\ref{in}, \ref{def}):
\begin{equation}\label{tab1}
\begin{tabular}{ c|| c| c| c| c| c}
\smaller[1.5]{$\hat{n}$} & \smaller[1.5]{$0$} & \smaller[1.5]{$1$} &\smaller[1.5]{$2$}& \smaller[1.5]{$3$} &\smaller[3]{\dots}\\ 
\hline
 \smaller[1.5]{} & \smaller[3]{$(0,0)$} & \smaller[3]{$(0,1)$, $(1,0)$, } & \smaller[3]{$(0,2)$, $(1,1)$, $(2,0)$ }& \smaller[3]{$(0,3)$,  $(1,2)$, $(2,1)$, $(3,0)$} & \smaller[3]{\dots}\\
 \hline
\smaller[1.5]{$\rho$}  & \smaller[1.5]{$1$} & \smaller[1.5]{$y^\alpha$, $y$} & \smaller[1.5]{$y^{2 \alpha}$, $y^{\alpha+1}$, $y^2$}& \smaller[1.5]{$y^{3 \alpha}$, $y^{2 \alpha+1}$, $y^{\alpha+2}$, $y^3$}& \smaller[3]{\dots} 
\end{tabular}
\end{equation}
where  $y \equiv e^{-\beta \omega_1 }$. 
The first, second and third row in (\ref{tab1}) show the eigenvalues of (resp.) $\hat{n}$, $(\hat{n}_1,\hat{n}_2)$ and $\rho$, with the prefactor $\xi$ is omitted; cf.~(\ref{in}). 
The unitary process (\ref{star}, \ref{coltrane}) permutes the eigenvalues of $\rho$.
Using (\ref{tab1}) one calculates averages of $\n$ and $\n_i=a_i^\dagger a_i$:
\begin{equation}\label{ex1}
\begin{split}
\langle \hat{n} \rangle &= \xi \left(1 y^\alpha +1 y + 2 y^{2 \alpha} + 2 y^{\alpha+1} + 2y^2 + \dots \right),\\
    \langle \hat{n}_1 \rangle &= \xi \left(0 y^\alpha +1 y + 0 y^{2 \alpha} + 1 y^{\alpha+1} + 2y^2 + \dots \right),\\
    \langle \hat{n}_2 \rangle &= \xi \left(1 y^\alpha +0 y + 2 y^{2 \alpha} + 1 y^{\alpha+1} + 0y^2 + \dots \right).
\end{split}
\end{equation}
The eigenvalues of $\rho$ in (\ref{tab1}) are organized in columns. Whenever the maximal element $y^{k\alpha}$ of $k$'th column is larger than the minimal element $y^l$ of $l$'th column ($l<k$), we interchange them and achieve some cooling. 
Formally, we should iterate till all elements in the third row are arranged in  descending order; cf.~(\ref{ohta}). Thus the optimal cooling increases the probability of eigenstates of $\hat{n}$ with lower photon number. 
Note from (\ref{tab1}, \ref{ex1}) that we can interchange elements within each column without changing $\Delta n$. 

When $\Delta n$ is fixed, the descending order of $\rho$'s eigenvalues in the final state yields simultaneously the minimum value of $\Delta n_1$ and the maximum value of $\Delta n_2$. This is because the eigenvalues of $\hat{n}_1$ ($\hat{n}_2$) in (\ref{tab1}) are arranged in ascending (descending) order. Eqs.~(\ref{cop2}, \ref{eff}) show that thereby also $\eta$ and $K$ reach their maximal values at the optimal $\Delta n$. The rule (\ref{ex1}) stays intact and can be used after permutations. 

The exact calculation of (\ref{coltrane}) is out of reach, since $\rho$ has an infinite number of eigenvalues. But we can develop a useful bound for it by focusing on permutations between nearest-neighbour columns. Define from (\ref{ratio}):
\BEA
\label{m2}
m \equiv \lceil\frac{\alpha}{1-\alpha}\rceil, 
\EEA
where $\lceil c \rceil$ is the smallest integer $\geq c$. Looking at (\ref{tab1}) it is seen that for $k\geq m$, the maximal element of the $(k+1)$'th column is larger than the minimal element of $k$'th column. 
Permuting them will contribute to $\Delta n$ calculated via (\ref{coltrane}). Likewise, for $k\geq m+2$, the next to maximal element of the $(k+1)$'th column is larger than the next to minimal element of $k$'th column. To visualize this situation consider a part of 
(\ref{tab1}) between columns $m+p$ and $m+p+1$ ($p\geq 0$):
\BEA
\begin{tabular}{ c|| c| c}
$\hat{n}$ & $m+p$ & $m+p+1$ \\ 
\hline
$\rho$  & 
$\dots$, $y^{\alpha+m+p-1}$, $y^{m+p}$ & $y^{(m+p+1)\alpha}$, $y^{(m+p)\alpha+1}$, $\dots$\\
\end{tabular}\nonumber\\
\label{tab2}
\EEA
where we omitted the second row of (\ref{tab1}).
Continuing this logic, we see that a new permutation appears for each even $p$, and that we can cover all nearest-neighbor permutations. Hence a bound [cf.~(\ref{in}, \ref{coltrane})]: 
\begin{gather}
0<-\Delta n_{\rm opt}
\equiv{\sum}_{k=0}^\infty (n_kr_k-n_k^{\uparrow}r_k^{\downarrow})\nonumber\\
\geq\xi {\sum}_{l=0}^\infty y^{l(\alpha+1)}\times
\sum_{k=m}^\infty (y^{(k+1)\alpha}-y^{k})
\nonumber\\
=\frac{(1-y)y^{\alpha(m+1)}
-(1-y^\alpha)y^m}{1-y^{\alpha+1}}.
\label{orb}
\end{gather}
According to (\ref{orb}), cooling is possible for any $0\leq \alpha<1$, i.e. (\ref{orb}) is positive and grows with $y=e^{-\beta\omega_1}$ changing from $0$ (at $y=0$) to $\frac{1-\alpha}{1+\alpha}$ at $y=1$. Appendix \ref{ap1} studies the optimal cooling numerically. In particular, it shows numerical plots for the optimal $K_{\rm opt}$ (COP) and $\eta_{\rm opt}$ (efficiency).

Now assume that $m$ given by (\ref{m2}) satisfies $m\gg 1$. Then the bound (\ref{orb}) gets small, and becomes nearly exact, since the relative error between $\Delta n_{\rm opt}$ and (\ref{orb}) scales as ${\cal O}(y^{2m})$. This estimate follows from the contribution of next to nearest-neighbor permutations and is confirmed in Appendix \ref{ap2}. We report here the limiting values of $K$ and $\eta$ only, which are obtained as described above [cf.~(\ref{cop2}, \ref{eff}, \ref{carnot})]:
\BEA
\label{gla}
&&\alpha\to 1: \quad K_{\rm opt} \to \infty,
\quad \eta_{\rm opt} \to 0,\\
&&\alpha\to 0: \quad K_{\rm opt} \to \infty,
\quad \eta_{\rm opt} \to 1,
\label{glad}
\EEA
where $\alpha=\omega_2/\omega_1\to 0$ in (\ref{glad}) is understood in the sense of a large $\omega_1$ and a small $\omega_2$. { It is also important to note that both $\Delta n_1$ and $\Delta n_2$ are functions of $\alpha$ and in the limit of $\alpha \to 1$ both tend to zero.}
In the last limits of (\ref{gla}, \ref{glad}) $\eta$ coincides with Otto bound. In both limits the energy costs of cooling are negligible: $K_{\rm opt} \to \infty$. In the more general case of large $\omega_1$  and fixed $\omega_2$, we studied $\eta_{\rm opt}$ and $K_{\rm opt}$ in Appendix \ref{ap3}.

\section{Feasible interaction Hamiltonian for cooling} 
\label{sec4}

How is a permutation unitary $U_{\rm opt}$ realized? This relates to one of major questions of quantum control; see e.g. \cite{rabitz}. Any Hamiltonian that is a polynomial of a fixed degree over $a_1$, $a_1^\dagger$, $a_2$ and $a_2^\dagger$ can be realized via sufficiently many linear operations plus a single-mode non-linearity \cite{lloyd}. However, realizing the permutation $U_{\rm opt}$ should be difficult in practice, since it refers to a Hamiltonian that is a highly non-linear over $a_1$, $a_1^\dagger$, $a_2$ and $a_2^\dagger$.

Now we focus on a feasible non-linear interaction and determine its cooling ability. The feasiblity comes at a cost: now cooling will be possible mostly next to nonlinear resonances: $\omega_2\gtrsim 2\omega_1$ or $2\omega_2\lesssim \omega_1$. This will also connect to the Manley-Rowe theorem, a known relation of nonlinear optics \cite{weiss,landau}.

{ The simplest $\chi^2$ nonlinear interactions can be realized in an anisotropic (e.g. crystalline) medium. Here the medium polarization $\vec{P}$ is quadratic in electric field $\vec{E}$ \cite{landau,new,hillery,walls}: $\vec{P} = \chi^{(1)}\vec{E}+\vec{E}\chi^{(2)}\vec{E}$, where $\chi^{(1)}$ and $\chi^{(2)}$ are susceptibilities. Neglecting the polarization degree of freedom for the electric field, its quantum operator representation is $\vec{E}\to a^{\dagger}+a$ \cite{hillery,walls}. Hence, the nonlinear interaction can be written as
\begin{equation}
\label{eq:inter}
    H_I = (a_1^{\dagger}+a_1)(a_2^{\dagger}+a_2)^2+(a_1^{\dagger}+a_1)^2(a_2^{\dagger}+a_2),
\end{equation}
with the full Hamiltonian of the system being
\begin{equation}
\label{full}
    H = \omega_1 a_1^{\dagger}a_1 + \omega_2 a_2^{\dagger}a_2 + gH_I = H_0 + gH_I
\end{equation}
where $g$ is the interaction constant.

Yet another scenario for (\ref{full}) is realized in the optomechanics. In addition to its applications in quantum technologies \cite{cavoptomech}, this field emerged as a potential basis for quantum gravity and foundations of quantum mechanics \cite{grav1,grav2}. In the optomechanical setting, the interaction between a laser and a mechanical oscillator is such that the resonance frequency $\omega_1(x)$ of the laser depends on the position $x$ of the mechanical oscillator. Hence their joint Hamiltonian reads: $H = \omega_1(x)a_1^\dagger a_1 + \omega_2a_2^\dagger a_2$ \cite{cavoptomech}. Here $a_1$ and $a_2$ are the annihilation operators for (resp.) the laser and the mechanical oscillator. Keeping up to the linear term of the Taylor expansion of $\omega_1(x)$ and using $x = a_2^\dagger + a_2$ we get 
\BEA
H = \omega_1 a_1^\dagger a_1 + \omega_2 a_2^\dagger a_2 + (\partial_x \omega_1) a_1^\dagger a_1 (a_2^\dagger + a_2),
\EEA
which closely relates to (\ref{full}).}

To employ (\ref{eq:inter}, \ref{full}) in (\ref{gulg}) we introduce the free Heisenberg interaction Hamiltonian $H_I(t)=e^{iH_0t} H_I e^{-iH_0t}$ and represent 
$\rho(t)={\rm e}^{-itH}\rho\, {\rm e}^{itH}$ in (\ref{star}) via chronological exponent
$\overleftarrow{\rm e}$:
\BEA
\label{kagan}
    \rho(t) = {\rm e}^{-i H_0 t} \widetilde{U} \rho\,\widetilde{U}^{\dagger}{\rm e}^{i H_0 t}, ~~
    \widetilde{U}=\overleftarrow{\rm e}^{
    -i\int_0^t ds\, gH_I(s)}.
\EEA
Now expand $\widetilde{U}$ into Dyson series
\BEA
    \widetilde{U}&=&1-ig
    \int_0^t ds H_I(s)\nonumber\\ 
    &-&g^2\int_0^t ds_1 \int_0^{s_1} ds_2 H_I(s_1) H_I(s_2)+ ...
\label{eq:dyson}
\EEA
Using $e^{iH_0 s}a_{k} e^{-iH_0 s}=  e^{-i\omega_{k}s}a_{k}$ ($k=1,2$) in $H_I(t)$, one can show that the order of magnitude estimate of
the $k$'th term in (\ref{eq:dyson}) reads
\begin{gather}
g^k\Omega^{-k}\,\sin^k\left( {\Omega t}/{2} \right),\\
\Omega={\rm min}[\omega_1, \omega_2,|2\omega_1-\omega_2|, |2\omega_2-\omega_1|].
\end{gather}
Thus, for a suitable $g$, $\omega_1$ and $\omega_2$ we can keep in (\ref{eq:dyson}) the first three terms. Within this weak-coupling approximation we calculated (\ref{gulg}) in Appendix \ref{ap4} showing that sufficiently large cooling $\Delta n<0$ is possible only for 
\BEA
\label{urk}
\omega_2\gtrsim 2\omega_1\quad {\rm or}\quad 
2\omega_2\lesssim \omega_1, 
\EEA
i.e. for two possible near-resonance conditions. Restricting ourselves with the latter case $\alpha\equiv\omega_2/\omega_1\lesssim 0.5$
we note that terms $a_1 a_2^{\dagger 2} + a_1^\dagger a_2^2$ in (\ref{eq:inter}) oscillate much slower than other terms. Hence within the rotating wave approximation we can take in (\ref{eq:inter}): 
\BEA
H_I  \simeq\overline{H}_I\equiv a_1 a_2^{\dagger 2} + a_1^\dagger a_2^2.
\label{grm}
\EEA
The approximation is studied in Appendix \ref{ap4},
where we also work out (\ref{eq:inter}).
Now $\overline{H}_I$ in (\ref{grm}) leads to an exact operator conservation:
\BEA
\label{botkin}
2\hat{n}_1+\hat{n}_2={\rm const},~~\hat{n}_k=a^\dagger_k a_k, ~~k=1,2.
\EEA
This conservation is the Manley-Rowe theorem for the considered nonlinear system \cite{weiss,landau}. The theorem does not generally hold for the complete interaction Hamiltonian (\ref{eq:inter}). However, the cooling necessitates $\alpha\lesssim 0.5$ (or $\alpha\gtrsim 2$) and is accompanied by an approximate conservation law (\ref{botkin}) (or $\hat{n}_1+2\hat{n}_2={\rm const}$). 
Using (\ref{grm}, \ref{botkin}) we get from (\ref{eq:dyson}, \ref{kagan}, \ref{gulg}) keeping there the first three terms only (the order of $g^2$): 
\begin{gather}
   \label{nur0}
    \Delta n_1 = \frac{8 g^2 \sin^2\left( \frac{(2 \omega _2-\omega _1)t}{2}\right)}{\left(2 \omega _2-\omega _1\right){}^2}
    \frac{\left(
    e^{\beta\omega _1}-e^{2 \beta\omega _2}
    \right)}{\left(e^{\beta\omega _1}-1\right)  \left(e^{\beta\omega _2}-1\right){}^2}, \\
   \Delta n_2 = -2\Delta n_1,\, \Delta n = -\Delta n_1,
   \label{nur}
\end{gather}
Hence the cooling at $\alpha\lesssim 0.5$ is described via 
$\eta = \frac{1}{3}$ and $K = \frac{1}{1-2\alpha}$; cf.~(\ref{cop2}, \ref{eff}). Once $\eta$ is finite and $K$ is large, we achieve cooling with a small energy cost.

{ Eq.~(\ref{nur0}) shows that a sizable cooling is achieved for sufficiently long times, because $\sin^2\left(\frac{(2 \omega _2-\omega _1)t}{2}\right)$ maximizes for ${|2 \omega _2-\omega _1|t}\sim {\pi}$, while $|2 \omega _2-\omega _1|$ is small; cf.~(\ref{urk}). This relation resembles the third law for the ordinary (energy) cooling, though more efforts are needed for its systematic investigation; e.g. we need a more complete understanding of the evolution generated by (\ref{full}).  }

\section{Summary} 

Our starting point was that linear transformations on boson modes (linear optics) increase the overall mean photon number, provided that the initial state is (generalized) diagonal; see (\ref{4}, \ref{44}) and (\ref{garu}). This unidirectional relation refers to the linear evolution, but applies for a wider set of initial states (\ref{4}, \ref{44}) than the second law does. Its similarities and differences with respect to the second law are discussed in section \ref{sl}. In its full generality this relation is formulated for the first time, though the literature was close to its formulation several times \cite{caves,karen}. Given that the lion's share of boson dynamics is linear, this general result will hold for a number of fields including optics, phononics {\it etc}. 
Importantly, we show explicitly that relation (\ref{garu}) connects to increasing the overall noise in the system (though its subsystems can get a noise reduction, as e.g. happen in squeezing \cite{garrison}). Hence we interpret
it as heating. 

{ It is interesting to ask how specifically the increase (\ref{garu}) of the overall mean photon number for initial states (\ref{4}, \ref{44}) relates to the second law. To answer this question, we studied the behavior of the Bose-entropy (\ref{uu1}) for linear dynamics and for the same class of initial states (\ref{4}, \ref{44}). The Bose-entropy is conditionally maximized at equilibrium, and it can change during unitary evolution in contrast to the (fine-grained) von Neumann entropy. We show in section \ref{entro} that for a subclass of linear evolution the Bose-entropy (\ref{uu1}) increases, and this increase also demands more restricted initial states (\ref{4}, \ref{44}, \ref{444}) than the validity of (\ref{garu}). A precise definition of this subclass relates to certain non-trivial problems in linear algebra. We thus confirm that for linear evolution the increase (\ref{garu}) of the overall mean photon number is a more general unidirectional relation than the second law. }

We show that the inverse of the heating in terms of the mean photon number (i.e. cooling) is possible within nonlinear (inter-mode) interactions. The cooling interpretation is not arbitrary and is characterized by efficiency and coefficient of performance (COP). { The former holds Otto's bound of the heat-engine efficiency (i.e. Carnot efficiency with temperatures replaced by frequencies)}. For the COP we anticipated, but so far did not identify, a general relation similar to Carnot's bound for the refrigeration COP \cite{armen2}.

We studied feasible nonlinear processes (e.g. $\chi^2$ \cite{landau,new,hillery,walls}) on two modes with different frequencies $\omega_1$ and $\omega_2$. Then the cooling in terms of the mean photon number happens (mostly) in the vicinity of nonlinear resonances. We also studied the optimal cooling, which is possible for any $\omega_1\not=\omega_2$, but is demanding from the viewpoint of dynamic realization. 

\begin{acknowledgments}
We are  grateful to Karen Hovhannisyan for important remarks and to David Petrosyan for discussions. This work was supported by SCS of Armenia, grant No. 20TTAT-QTa003.  
\end{acknowledgments}

\bibliography{apssamp}

\clearpage

\appendix

\section{ Examples and counter-examples for inequality (\ref{ds})}
\label{ap0}

We are given any $2\times 2$ matrix
\BEA
\begin{pmatrix} a & b \\ c & d \end{pmatrix}.
\label{aa1}
\EEA
with non-negative elements and
\BEA
a+b\geq 1, \quad c+d \geq 1, \quad a+c \geq 1, \quad b+d \geq 1.
\EEA
Define 
\BEA
c={\rm min}[a,b,c,d]
\EEA
and note that only $c<1$ is non-trivial, since otherwise (\ref{ds}) holds for (\ref{aa1}) and 
any double-stochastic matrix. Now 
\BEA
\begin{pmatrix} a & b \\ c & d \end{pmatrix}
\geq  \begin{pmatrix} 1-c & c \\ c & 1-c \end{pmatrix}=\Theta,
\EEA
where the latter matrix is double-stochastic. 
Thus (\ref{ds}) holds for $N=2$.

The simplest counter-examples for (\ref{ds}) at $N=3$ is the following matrix with non-negative elements:
\BEA
\begin{pmatrix} 
a_{11} & a_{12} & a_{13} \\ 
0      & a_{22} & a_{23} \\
0      & a_{32} & a_{33} \\
\end{pmatrix},
\EEA
where $a_{21}=a_{31}=0$. We assume
\BEA
{\sum}_{i=1}^3a_{ik}\geq 1, \quad {\sum}_{k=1}^3a_{ik}\geq 1, \quad  
\EEA 

and additionally
\BEA
\label{condlast}
a_{22}+a_{32}<1.
\EEA

If a double-stochastic matrix holding (\ref{ds}) exists, then we have
\BEA
\begin{pmatrix} 
a_{11} & a_{12} & a_{13} \\ 
0      & a_{22} & a_{23} \\
0      & a_{32} & a_{33} \\
\end{pmatrix}\geq
\begin{pmatrix} 
1      & 0           & 0 \\ 
0      & \Theta_{22} & \Theta_{23} \\
0      & \Theta_{32} & \Theta_{33} \\
\end{pmatrix}.
\EEA
Now the latter matrix cannot be double-stochastic and hold (\ref{ds}), as the condition (\ref{condlast})  
is violated, if we take in (\ref{lolo}) ${\cal I}={2,3}$ 
and ${\cal J}={1,2}$.

\section{ Numerical results for the optimal cooling}
\label{ap1}

Recall our discussion after (\ref{tab1}) of the main text. There we explained that the optimal cooling|with respect to all involved quantities $\Delta n_{\rm opt}$ (photon number difference), $K_{\rm opt}$ (COP or coefficient of performance) and $\eta_{\rm opt}$ (efficiency)|is achieved once all eigenvalues of the final density matrix are arranged in the descending order; see the third row in (\ref{tab1}) of the main text. Numerically, this means that we need to take a sufficiently long but a finite sequence of eigenvalues (starting from the largest one) and ensure that the results are stable with respect to increasing the length of this block. 

\begin{figure}[h!]
    \centering
    \includegraphics[scale=0.37]{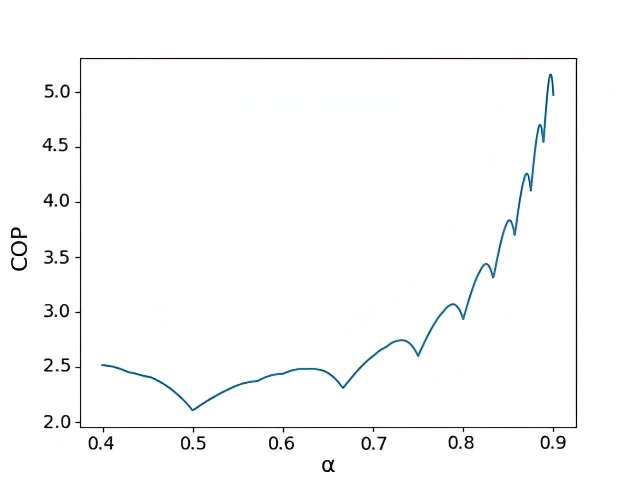}
    \caption{The optimal coefficient of performance (COP) $K_{\rm opt}$ versus $\alpha=\omega_2/\omega_1<1$ for the optimal cooling. Here $y^{\alpha}=e^{-\beta\omega_2} = 0.6$ and numerical calculations are done up to the block number $300$; see (\ref{tab1}) of the main text.}
\label{smm1}
\end{figure}

\begin{figure}[h!]
    \centering
    \includegraphics[scale=0.37]{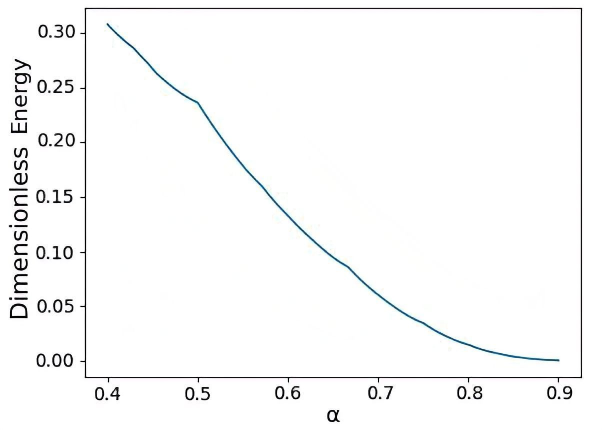}
    \caption{The same as in Fig.~\ref{smm1}, but for the optimal dimensionless energy cost $\Delta n_1 + \alpha\Delta n_2$ versus $\alpha$ defined via (\ref{ratio}).}
\label{smm2}
\end{figure}

\begin{figure}[h!]
    \centering
    \includegraphics[scale=0.37]{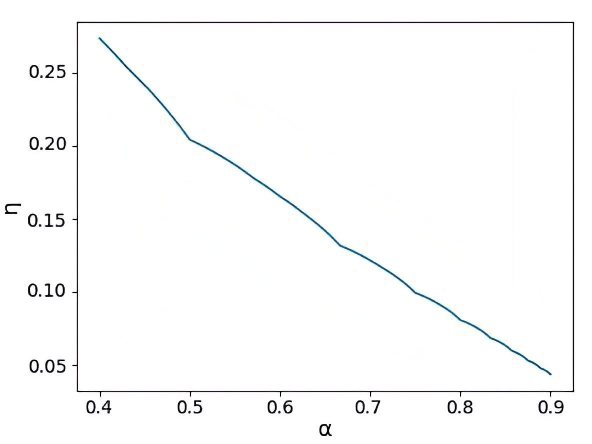}
    \caption{The same as in Fig.~\ref{smm1}, but
    for the efficiency $\eta_{\rm opt}$ versus $\alpha$.}
\label{smm3}
\end{figure}

Our numerical results are shown in Figs.~\ref{smm1}, \ref{smm2} and \ref{smm3}. First, recall that in $K = -\frac{\Delta n_1 + \Delta n_2}{\Delta n_1 + \alpha\Delta n_2}$, the achieved photon number decrease $\Delta n=\Delta n_1 + \Delta n_2<0$ is divided over the dimensionless energy cost $\Delta n_1 + \alpha\Delta n_2$; cf.~(\ref{cop2}) of the main text. It is seen from Fig.~\ref{smm1} that $K_{\rm opt}$ as a function of $\alpha$ [cf.~(\ref{ratio})] has (singular) local minima at points $\alpha = \frac{k}{k+1}$ where $k\in \mathbb{N}$ is an integer. We checked that these local minima of $K$ come mostly from the singular behavior of the energy cost $\Delta n_1 + \alpha\Delta n_2$; see Fig.~\ref{smm2}. Now $\Delta n$ (not shown in figures) shows weak singularities at those points $\alpha = \frac{k}{k+1}$, but these singularities are much weaker than those of the energy cost $\Delta n_1 + \alpha\Delta n_2$. 

The origin of these singularities for $K_{\rm opt}$ (and $\Delta n_1 + \alpha\Delta n_2$) can be clarified as follows. Recall that [cf.~(\ref{m2})]
\BEA
\label{mmm}
m \equiv \left\lceil\frac{\alpha}{1-\alpha}\right
\rceil,
\EEA
refers to the the group of eigenvalues of the initial state $\rho$ starting from which the eigenvalues of $\rho$ are not arranged in the descending order; see the discussion after (\ref{tab1}) of the main text. At points $\alpha = \frac{k}{k+1}$ the index of the block from which the permutations start undergoes a jump discontinuity of increasing by one. 

Fig.~\ref{smm3} presents the numerical behavior of $\eta_{\rm opt}$ as a function of $\alpha$. It is seen that $\eta_{\rm opt}$ also shows singularities at $\alpha = \frac{k}{k+1}$, though these singularities are weaker than those for $K_{\rm opt}$; cf.~Fig.~\ref{smm1}. In particular,  
these singularities do not change the monotonous behavior of $\eta_{\rm opt}$ as a function of $\alpha$.

\section{ Asymptotic results for the optimal cooling: The limit $\alpha \to 1$}
\label{ap2}

\begin{figure}[h]
    \centering
    \includegraphics[scale=0.37]{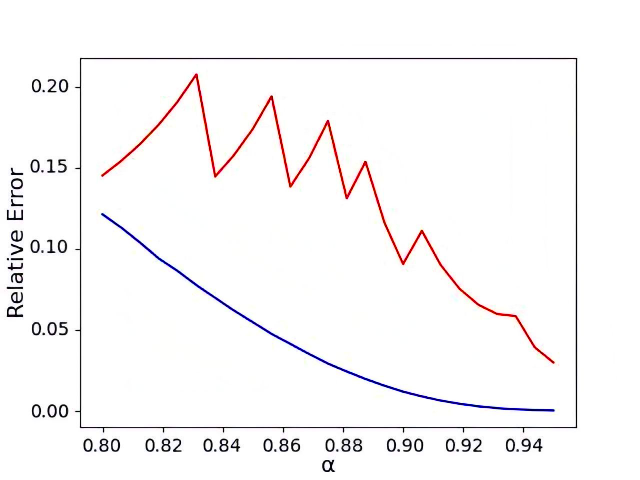}
    \caption{Here we calculated $\Delta n$ in numerically exact way by directly arranging all eigenvalues of the final density matrix $\rho(t)$ in the descending order; cf.~(\ref{tab1}) of the main text. We took $y^{\alpha}=e^{-\beta\omega_2} = 0.8$ and numerical calculations were done up to block number 100 which is greater than $5m$ [cf.~(\ref{mmm})] for all those $\alpha$ values included in the graph. This quantity was denoted by $\Delta n_{\rm exact}$. We denote via $\Delta n_{\rm nn}$ the neasrest neighbor approximation given by (\ref{orb}) of the main text. \\ 
    Blue curve: the relative error $\frac{|\Delta n_{\rm exact}-\Delta n_{\rm nn}|}{\Delta n_{\rm nn}}$. Red curve: $|\frac{y^{2m(\alpha)}}{\Delta n_{\rm nn}}|$. This curve is kinked, because so is $m(\alpha)$; see (\ref{mmm}). It is seen that the relative error is well within the announced range $|\frac{{\cal O}( y^{2m(\alpha)})}{\Delta n_{\rm nn}}|$.
    }
\label{sm1}
\end{figure}

Eq.~(\ref{orb}) of the main text provides the nearest-neighbour approximation for $\Delta n$. There we also indicated that (\ref{orb}) of the main text becomes close to its exact value whenever $m$ defined via (\ref{mmm}) is sufficiently large, or, equivalently $\alpha \to 1$. The precision of this approximation relates to the necessity of next-nearest-neighbour permutations. The largest value of $p$ in (\ref{tab2}) of the main text, where such permutations are necessary can be estimated from the following diagram:
\BEA
\begin{tabular}{ c|| c| c |c|}
$\hat{n}$ & $2m$ & $2m+1$ & $2m+2$ \\ 
\hline
$\rho$  & 
$\dots$, $y^{2m}$ & $\dots$ & $y^{(2m+2)\alpha}$, $\dots$\\
\end{tabular}
\label{tab7}
\EEA
Now note from (\ref{mmm}) that $y^{2m}<y^{(2m+2)\alpha}$, i.e. a next-nearest-neighbor permutation is necessary.
Hence the contribution from next-nearest-neighbor permutation scales as ${\cal O}(y^{2m})$, and for $m\gg 1$ this is smaller than what was retained in (\ref{orb}) of the main text. This estimate is crude, since it did not account for permutations that already occurred (within the nearest-neighbor approach) between the columns $2m$ and $2m+1$. However, it is sufficient for our purposes. Indeed, Fig.~\ref{sm1} shows the relative error of numerically exact calculation of $\Delta n$ and compares it with (\ref{orb}) of the main text showing that it is well within the above bound 
${\cal O}(y^{2m})$.

\subsection{ COP in the limit $\alpha \to 1$}
\label{ap2.1}

For studying COP $K$, we can write the mean changes of $\n_1$ and $\n_2$ in the approximation of  nearest-neighbor permutations [cf.~(\ref{tab2}) of the main text]:
\begin{equation}
\begin{split}
    \Delta n_1 &= \xi{\sum}_{i=\tilde{n}}^{\infty} (y^{\alpha(i+1)}-y^{i}){\sum}_{j=0}^{\infty}y^{j(1+\alpha)}(i+j)\\
    \Delta n_2 &= -\xi{\sum}_{i=\tilde{n}}^{\infty} (y^{\alpha(i+1)}-y^{i}){\sum}_{j=0}^{\infty}y^{j(1+\alpha)}(i+j+1),
\end{split}
\label{otark}
\end{equation}
where $\xi=(1-e^{-\beta\omega_1}) 
    (1-e^{-\beta\omega_2})$ is the normalization factor; cf.~(\ref{in}) of the main text. 

Note that for obtaining (\ref{otark}) we do not make any permutation within columns with the same eigenvalue of $\hat{n}$; cf.~(\ref{tab1}, \ref{tab3}) of the main text. Doing such permutations will make the estimates in (\ref{otark}) closer to the minimal value of $\Delta n_1$ and the maximal value of $\Delta n_2$ (both for a fixed $\Delta n$). Hence (\ref{otark}) suffices for bounding $K$ from below:
\begin{equation}
    K \geq \frac{1}{(1-\alpha)(\frac{y^{1+\alpha}}{1-y^{1+\alpha}}+\frac{1}{1-y^\alpha})} \implies \lim_{\alpha\to1} K \to \infty,
\end{equation}
which is also observed numerically.

\begin{figure}[h]
    \centering
    \includegraphics[scale=0.37]{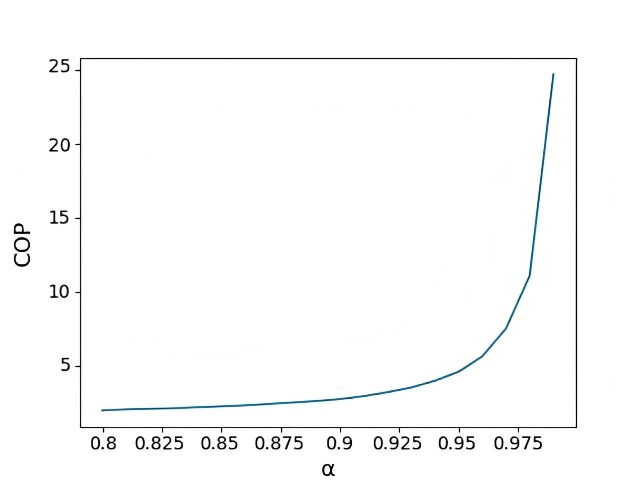}
    \caption{Numerical results for the coefficient of performance (COP) $K$. Here $y^{\alpha} = 0.8$ and numerical calculations are done up to block number $5m$ for all those $\alpha$ values included in the graph.}
    \label{cop3}
\end{figure}

\section{ Asymptotic results for the optimal cooling: The limit $\alpha \to 0$}
\label{ap3}

\subsection{ Error estimation}
\label{ap3.1}

For $\alpha$ finite and sufficiently close to $0$, the action of an optimal unitary results in (\ref{tab31}) 
\begin{equation}\label{tab31}
\begin{split}
&\begin{tabular}{ c| c| c| c| c| c|}
$N$ & $0$ & $1$ & $\dots$ & $a'$ & $\dots$\\ 
\hline
$R$ & $1$ & $y^\alpha$, $y^{2 \alpha}$ & $\dots$& $\dots$, $y^{m_1 \alpha}$ , $y$, $y^{(m_1+1)\alpha}$, $\dots$ & $\dots$\\  
\end{tabular}\\&
\begin{tabular}{| c| c}
 $a'+i$  & $\dots$ \\ 
\hline
$\dots$, $y^{(m_2-1) \alpha}$, $y^2$, $y^{(m_2+1)\alpha}$ $y^{(m_2+2)\alpha}$, $\dots$ & $\dots$ \\  
\end{tabular},
\end{split}
\end{equation}
where $m_1 = \lfloor 1/\alpha \rfloor$, $m_2 = \lfloor 2/\alpha \rfloor \geq 2m_1$, $a'$ is determined from 
\BEA
a'(a'+1)/2 \leq m_1 \leq (a'+1)(a'+2)/2
\EEA
and $i$ from
\begin{equation} \label{i1}
(a'+i)(a'+i+1)/2 \leq m_2 \leq (a'+i+1)(a'+i+2)/2.
\end{equation}
We see that $i < 3a'$. Now we show that for the calculation of averages of photon numbers one can use 
\begin{equation}\label{tab3}
\begin{tabular}{ c| c| c| c| c| c}
$N$ & $0$ & $1$ & $2$& $3$ & $\dots$\\ 
\hline
$R$  & $1$ & $y^\alpha$, $y^{2 \alpha}$ & $y^{3 \alpha}$, $y^{4 \alpha}$, $y^{5 \alpha}$& $y^{6 \alpha}$, $y^{7\alpha}$, $y^{8 \alpha}$, $y^{9 \alpha}$& $\dots$ \\  
\end{tabular}.
\end{equation}
instead of (\ref{tab31}), as in the limit $\alpha \to 0$ corresponding error terms vanish.
We denote by $n_1^{(0)}$ and $n_1^{(*)}$ the average $\n_1$ calculated with (\ref{tab31}) and (\ref{tab3}) respectively, and by $\Delta n_{1e}$ the error term $n_1^{(0)} - n_1^{(*)}$. Firstly, we write the contribution from the $a'^{\text{th}}$ block to the error term 
\begin{equation}
\begin{split}
    \Delta n^{a'}_{1e}&= (y-y^{(m_1+1)\alpha})(m_1+1-u) \\
    &+ (y^{(m_1+1)\alpha}-y^{(m_1+2)\alpha})(m_1+2-u)+ \dots \\& + (y^{(u+a'-1)\alpha}-y^{(u+a')\alpha}) a',
\end{split}
\end{equation}
where $u = a'(a'+1)/2$.
$\Delta n^{a'}_{1e}$ can be estimated from above 
\begin{equation}
\begin{split}
    \Delta n^{(a')}_{1e}&\leq (y-y^{(m_1+1)\alpha}) a'+ (y^{(m_1+1)\alpha}-y^{(m_1+2)\alpha}) a'+ \\ &+ \dots  + (y^{(u+a'-1)\alpha}-y^{(u+a')\alpha}) a' = \\
    &= (y-y^{(u+a')\alpha}) a'.
\end{split}
\end{equation}
Similarly, one can estimate the contribution from $(a'+1)^{\text{th}}$ block
\begin{equation}
    \Delta n^{(a'+1)}_{1e} \leq (y^{(u+a')\alpha}-y^{(u+2a'+2)\alpha}) (a'+1).
\end{equation}
Summing up all contributions we get 
\begin{equation}\label{errf1}
\begin{split}
    &\Delta n_{1e} \leq \big[y + y^{(u+a')\alpha} + y^{(u+2a'+2)\alpha} + y^{(u+3a'+5)\alpha}  + \dots \big] \\&+
    \big [(y^2- y^{(m_2+1)\alpha})(a'+i) + (y^3- y^{(m_3+1)\alpha})(a'+i') \\&+ (y^4- y^{(m_4+1)\alpha})(a'+i'')+\dots \big ],
\end{split}
\end{equation}
where $m_3=\lfloor3/\alpha\rfloor\geq 3m_1$, $m_4=\lfloor4/\alpha\rfloor\geq 4m_1$. $i'$ and $i''$ in (\ref{errf1}) are determined from conditions similar to (\ref{i1}):
\begin{equation}
    \begin{split}
        &\frac{(a'+i')(a'+i'+1)}{2} \leq m_3 \leq \frac{(a'+i'+1)(a'+i'+2)}{2}, \\
        &\frac{(a^{'}+i^{''})(a^{'}+i^{''}+1)}{2} \leq m_4 \leq \frac{(a^{'}+i^{''}+1)(a^{'}+i^{''}+2)}{2}
    \end{split}
\end{equation}
and result in $i' < 4a'$ and $i'' < 5a'$ Now, we can estimate (\ref{errf1}) further 
\begin{equation}
\begin{split}
        &\Delta n_{1e} \leq \big[y + y^{2 } + y^{3 } + y^{4 }  + \dots \big] \\&+
    \big [y^{2}3 a' + y^{3 } 4 a'+ y^{4} 5 a'+\dots \big ]\leq\\ 
    &\leq \frac{y(1+y)}{1-y} + a' \frac{y \ln {y} }{(1-y)^2}.
\end{split}
\end{equation}
Now, note that in the limit $\alpha \to 0$, which is $\omega_1 \to \infty$, $a'$ goes to infinity as $\sqrt{\frac{\omega_1}{\omega_2}}$. As, $ 0 \leq \Delta n_{1e} \leq \frac{y(1+y)}{1-y} + a' \frac{y \ln {y} }{(1-y)^2}$ we conclude that 
(remember that $y = e^{-\omega_1\beta}$)
\begin{equation}
        \lim_{\alpha \to 0}\Delta n_{1e} \to 0.
\end{equation}

\subsection{ Asymptotic expressions and their integral representations}
\label{ap3.2}

\begin{figure}[h]
    \centering
    \includegraphics[scale=0.38]{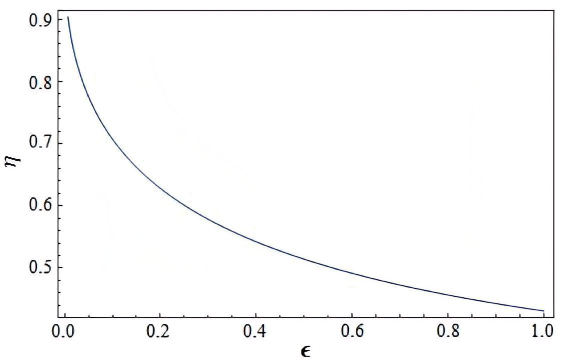}
    \caption{Numerically calculated efficiency ($\eta$) versus $\epsilon$ in the limiting case $\omega_1 \gg \omega_2$ using  (\ref{eq:transformed1}) and (\ref{eq:transformed2}). Here, we set $\beta \omega_1 = 10$ and the smallest value of $\epsilon$ is 0.007. }
\label{eta}
\end{figure}

\begin{figure}[h]
    \centering
    \includegraphics[scale=0.38]{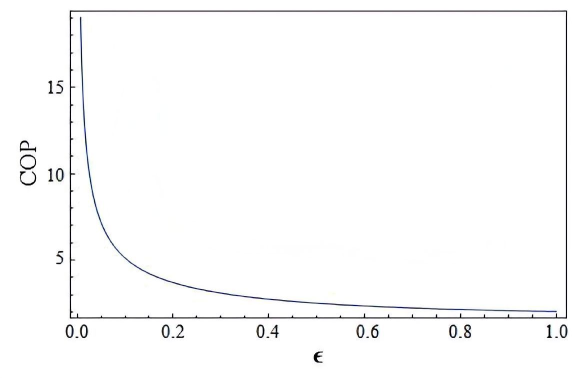}
    \caption{The same as in Fig.~\ref{eta} but for COP (K).}
\label{cop}
\end{figure}

In our further calculations we use (\ref{tab3}). 
Using the same procedure as in (\ref{ex1}), we find from (\ref{tab3}) the following expressions for $\Delta n_{1,2}$ ($\epsilon = \alpha \beta \omega_1$) 
\begin{equation}\label{eq:a0ns}
\begin{split}
    &\Delta n_1 =  \xi
    \sum_{a=0}^\infty e^{-\epsilon \frac{a(a+1)}{2}} \sum_{b=0}^a e^{-\epsilon  b } b - \xi_1 \sum_a  a e^{-\beta \omega_1 a} \\
    & =\xi \sum_{a=0}^\infty \frac{e^{2 \epsilon}} {(e^\epsilon -1)^2} e^{-\left(\frac{a}{2}+1\right) (a+1) \epsilon } \left(a \left(e^{-\epsilon }-1\right)+e^{a \epsilon }-1\right)\\& -\frac{1}{e^{\beta \omega_1}-1}, \\
    &\Delta n_2 =  \xi
    \sum_{a=0}^\infty e^{-\epsilon \frac{a(a+1)}{2}} \sum_{b=0}^a e^{-\epsilon (a -b) } b - \xi_2 \sum_a a e^{- \epsilon  a} \\
    &=\xi\sum_{a=0}^\infty \frac{ e^{\epsilon}} {(e^\epsilon-1)^2} e^{- \frac{a(a+1)}{2}\epsilon} (a (e^{\epsilon }-1 )+e^{-a \epsilon}-1)\\&- \frac{1}{e^{\epsilon}-1},
\end{split}    
\end{equation}
where we defined $\xi_1=1-e^{-\beta\omega_1}$ and $\xi_2=1-e^{-\beta\omega_2}$; hence $\xi=\xi_1\xi_2$. Before studying (\ref{eq:a0ns}) numerically, we apply  Hubbard-Stratonovich transformation: 
\BEA
e^ {-\frac{a^{2}}{2} \epsilon }=\sqrt{\frac{1}{2 \pi \epsilon}} \int_{-\infty}^{\infty} d v e^{-\frac{v^{2}}{2 \epsilon}-i a v},
\EEA
for faster and more accurate calculations:
\begin{equation}\label{eq:transformed1}
\begin{split}
\\&\sum_{a=0}^\infty e^{-\left(\frac{a}{2}+1\right) (a+1) \epsilon } \left(a \left(e^{-\epsilon }-1\right)+e^{a \epsilon }-1\right) \\& = 
\sqrt{\frac{1}{2 \pi \epsilon}} e^{-\epsilon}\int_{-\infty}^\infty dv e^{-\frac{v^{2}}{2 \epsilon}}\bigg( (e^{-\epsilon}-1) \frac{e^{-(i v+\frac{3}{2} \epsilon)}}{(1-e^{-(i v+\frac{3}{2} \epsilon)})^2} \\&+  \frac{1}{(1-e^{-(i v+\frac{1}{2} \epsilon)})} -  \frac{1}{(1-e^{-(i v+\frac{3}{2} \epsilon)})}\bigg),
\end{split}    
\end{equation}
\begin{equation}\label{eq:transformed2}
\begin{split}
\\&\sum_{a=0}^\infty e^{- \frac{a(a+1)}{2}\epsilon} (a (e^{\epsilon }-1 )+e^{-a \epsilon}-1) \\& = 
\sqrt{\frac{1}{2 \pi \epsilon}}   \int_{-\infty}^\infty dv e^{-\frac{v^{2}}{2 \epsilon}} \bigg((e^{\epsilon}-1)\frac{e^{-(i v+\frac{1}{2} \epsilon)}}{(1-e^{-(i v+\frac{1}{2} \epsilon)})^2} \\&+  \frac{1}{(1-e^{-(i v+\frac{3}{2} \epsilon)})} - \frac{1}{(1-e^{-(i v+\frac{1}{2} \epsilon)})}\bigg).
\end{split}    
\end{equation}
The results of numerical calculations for $\eta$ and $K$ are depicted in Fig.~\ref{eta} and Fig.~\ref{cop}. As seen from figures, $ K \to \infty $ and $\eta \to 1$ in the limit $\epsilon \ll 1$. Below, we show analytically, that indeed, $K$ and $\eta$ reach these limits.

\subsection{ Asymptotic results via the Euler-Maclaurin formula}
\label{ap3.3}

To study the asymptotics of $\eta$ and $K$ in the limit $\epsilon \ll 1$  we apply the Euler-Maclaurin formula  for the sums in (\ref{eq:a0ns})
\begin{equation}\label{fa}
\begin{split}
    &\sum_{a=0}^\infty \frac{e^{2 \epsilon}} {(e^\epsilon -1)^2} e^{-\left(\frac{a}{2}+1\right) (a+1) \epsilon } \left(a \left(e^{-\epsilon }-1\right)+e^{a \epsilon }-1\right) \\& \equiv \sum_{a=0}^\infty f_1(a) = \int_0^\infty dx f_1(x) + I_1 = S_1 + I_1
\end{split}
\end{equation}
where 
\begin{equation}
\begin{split}
I_1 &= \frac{f_1(0)+f_1(\infty)}{2} \\&+\sum_{k=1}^{\left\lfloor\frac{p}{2}\right\rfloor} \frac{B_{2 k}}{(2 k) !}\left(f_1^{(2 k-1)}(\infty)-f_1^{(2 k-1)}(0)\right)+R_{p},
\end{split}
\end{equation}

\begin{equation}
    R_p \leq \frac{2 \bm{\zeta}(p)}{(2 \pi)^p} \int_0^\infty dx |f_1^{(p)}(x)|, 
\end{equation}
$B_{2 k}$ are Bernoulli numbers, $\bm{\zeta}(p)$ is the Riemann's zeta function  
and $f^{(p)}(x)$ is the $p^\text{th}$ order differential. $p$ in (\ref{fa}) takes different integer values $p\geq 2$ and we use $p=2$, because this is the simplest case amenable to estimates.  
Similarly, 
 \begin{equation}\label{ga}
\begin{split}
    &\sum_{a=0}^\infty \frac{ e^{\epsilon}} {(e^\epsilon-1)^2} e^{- \frac{a(a+1)}{2}\epsilon} (a (e^{\epsilon }-1 )+e^{-a \epsilon}-1) \equiv  \sum_{a=0}^\infty f_2(a) \\&= \int_0^\infty dx f_2(x) + I_2 = S_2 + I_2,
\end{split}
\end{equation}
The leading diverging terms in (\ref{fa}) and (\ref{ga}) when $\epsilon \to 0$ are $S_1$ and $S_2$ and we omit $I_1$ and $I_2$.  
Using (\ref{fa}) and (\ref{ga}) for the efficiency and COP we get the following relations 
\begin{equation}
\begin{split}
   K &\approx -\frac{\xi S_1 - n_{1i}+\xi S_2- n_{2i}}{\xi S_1- n_{1i} + \alpha (\xi S_2-n_{2i})}, \\ \eta &\approx- \frac{\xi S_1-n_{1i}+\xi S_2- n_{2i}}{\xi S_1 - n_{1i}-\xi S_2  + n_{2i}}, 
\end{split}
\end{equation}
where $n_{1i}$ and $n_{2i}$ are initial average occupation numbers. The limits $\lim_{\alpha\to 0} \xi S_{1,2} / n_{2i}$ can be studied analytically, and we get $$\lim_{\alpha\to 0} \xi S_{1,2}/ n_{2i}  = 0.$$ Thus, for the $K_{\rm opt}$ and $\eta_{\rm opt}$ we obtain  
\BEA
K_{\rm opt} \to \infty,
\quad \eta_{\rm opt} \to 1.
\EEA

\section{ Perturbative treatment of the full nonlinear Hamiltonian}
\label{ap4}

Let us return to the full|i.e. without the rotating-wave approximation|nonlinear Hamiltonian given by (\ref{eq:inter}) of the main text: 
\BEA
\label{inter}
    H_I = (a_1^{\dagger}+a_1)(a_2^{\dagger}+a_2)^2+(a_1^{\dagger}+a_1)^2(a_2^{\dagger}+a_2).
\EEA
See (\ref{full}) of the main text for the complete Hamiltonian. Here we shall employ (\ref{inter}) in the second-order of Dyson's series given by (\ref{eq:dyson}) of the main text; see in this context (\ref{kagan}) of the main text. For simplicity we shall scale out the factor $\beta$, i.e. we denote $\beta g \to  g$, $\beta\omega_{1,2} \to \omega_{1,2}$ and $t/\beta \to t$. 

Using (\ref{eq:dyson}, \ref{kagan}) of the main text we get
\begin{equation}
\begin{split}\label{eq:dysoncooling}
    &{\rm tr}(\rho(t)\hat{n} - \rho(0)\hat{n}) = \mathcal{O}(g^3) + g^2\times\\
    &\times\bigg[\int_0^t ds H_I(s)\hat{n}\int_0^t ds H_I(s) - \\ &-  \int_0^t ds_1 \int_0^{s_1} ds_2 H_I(s_1) H_I(s_2)\hat{n} -\\&-  \hat{n}\int_0^t ds_1 \int_0^{s_1} ds_2 H_I(s_2) H_I(s_1)\bigg].
\end{split}
\end{equation}
Formally the same equation holds for $\hat{n}_k=a_k^\dagger a_k$, where $k=1,2$ and 
$\hat{n}=\hat{n}_1+\hat{n}_2$. 

Substituting (\ref{inter}) into (\ref{eq:dysoncooling}) we get
\begin{equation}
\label{eq:fullcooling}
\begin{split}
    &\Delta n_k ={\rm tr}\left(\rho(t)\hat{n}_k - \rho(0)\hat{n}_k\right)=\\
    = &g^2\bigg[ A_k\Phi(\omega_1+2\omega_2)+B_k\Phi(\omega_1-2\omega_2)+C_k\Phi(\omega_1)\bigg]+\\
    +&g^2 \bigg[D_k\Phi(\omega_2+2\omega_1)+E_k\Phi(\omega_2-2\omega_1)+F_k\Phi(\omega_2)\bigg],
\end{split}
\end{equation}
where $k=1,2$,
\begin{equation}
    \Phi(x) \equiv\frac{ 4\sin^2{(\frac{1}{2}xt)}}{x^2},
\end{equation}
\begin{equation}\label{eq:fullcoolingtermcoefs}
\begin{split}
    &A_1 = \frac{2 \left(e^{\omega_1+2\omega_2}-1\right)}{\left(e^{\omega_1}-1\right) \left(e^{\omega_2}-1\right)^2}, \quad
    A_2 = \frac{4 \left(e^{\omega_1+2\omega_2}-1\right)}{\left(e^{\omega_1}-1\right) \left(e^{\omega_2}-1\right)^2},\\
    &B_1 = \frac{2 \left(e^{\omega_1}-e^{2\omega_2}\right)}{\left(e^{\omega_1}-1\right) \left(e^{\omega_2}-1\right)^2}, \quad
    B_2 = \frac{-4 \left(e^{\omega_1}-e^{2\omega_2}\right)}{\left(e^{\omega_1}-1\right) \left(e^{\omega_2}-1\right)^2},\\
    &C_1 = \frac{4 e^{\omega_2}}{\left(e^{\omega_2}-1\right)^2}, \quad\qquad\qquad
    C_2 = 0.
\end{split}
\end{equation}
Now $D_1$, $E_1$ and $F_1$ are obtained from (resp.) $A_2$, $B_2$ and $C_2$ upon swapping 
$\omega_1$ and $\omega_2$. Likewise, $D_2$, $E_2$ and $F_2$ are obtained from (resp.) $A_1$, $B_1$ and $C_1$ upon swapping $\omega_1$ and $\omega_2$.

\begin{figure}[h]
    \centering
    \includegraphics[scale=0.37]{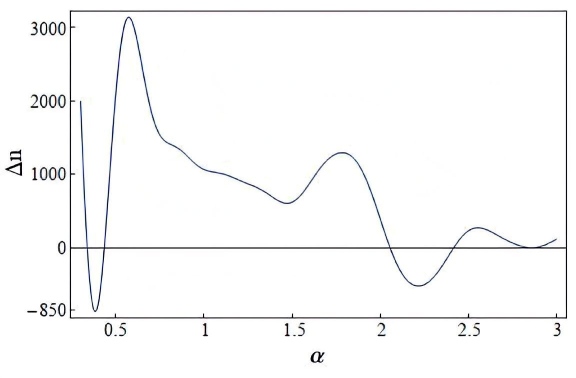}
    \caption{The photon  number difference $\Delta n$ obtained from (\ref{eq:fullcooling}) and (\ref{eq:fullcoolingtermcoefs}) for $\omega_1 = 0.35$ and $t = 10\pi$, where $\alpha=\omega_2/\omega_1$. It is seen that near the resonating frequencies $\alpha \lesssim 0.5$ and $\alpha \gtrsim 2$ the interaction Hamiltonian (\ref{inter}) results in cooling. We see that $\Delta n>0$ (no cooling) for other values of $\alpha$.}
    \label{resonance_cooling}
\end{figure}

\begin{figure}[h]
    \centering
    \includegraphics[scale=0.37]{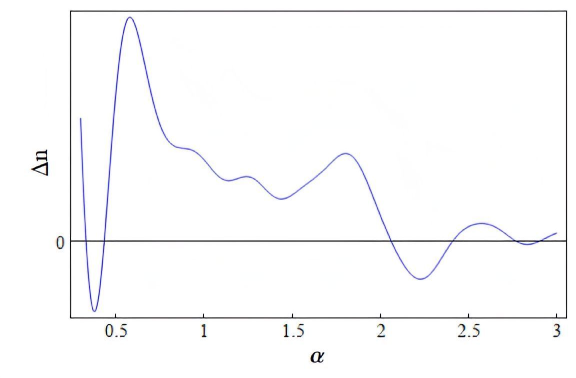}
    \caption{The photon  number difference $\Delta n$ obtained from (\ref{eq:fullcooling}) and (\ref{eq:fullcoolingtermcoefs}) for $\omega_1 = 0.6$ and $t = 6\pi$, where $\alpha=\omega_2/\omega_1$. We see that at the non resonance point $\alpha\approx2.8$ there are small values of $\Delta n < 0$ (cooling). Hence, although very small, cooling can also be achieved far from the resonance frequencies.}
    \label{resonance_cooling2}
\end{figure}

For a representative pair of frequencies $\omega_1$ and $\omega_2$, Fig.~\ref{resonance_cooling}
demonstrates to which extent $\Delta n=\Delta n_1+\Delta n_2$ calculated via (\ref{eq:fullcooling}) predicts cooling, i.e. $\Delta n<0$. As announced in the main text, cooling happens in near-resonance conditions $\omega_2\gtrsim 2\omega_1$ or $2\omega_2\lesssim \omega_1$, which is seen in Fig.~\ref{resonance_cooling}; see also Fig.~\ref{resonance_cooling2} for additional information. 

Now the essence of rotating-wave approximation in (\ref{eq:fullcooling}) is that e.g. for $|2\omega_2-\omega_1|\ll {\rm min}[\omega_1, \omega_2,|2\omega_1-\omega_2|]$, we can take $\Phi(\omega_1-2\omega_2)$ in (\ref{eq:fullcooling}) much larger than other terms. This reverts to (\ref{nur}) of the main text. 

\subsection{ Estimation of the higher-order terms in Dyson's series }
\label{ap4.1}

Using (\ref{eq:dyson}) one can show that the terms $\mathcal{O}(g^{l})$ in Dyson's series
(cf.~(\ref{eq:dysoncooling}) of the main text 
are based on the following structure:
\begin{equation}
    \begin{split}
        &\prod_{i=1}^{k}g \int_{0}^{s_{i-1}} ds_i H_I(s_i) \times \rho(0) \times \\
        &\prod_{i=1}^{k'}g \int_{0}^{s_{i-1}} ds_i H_I(s_{k'-i+1}) \hat n,
    \end{split}
    \label{eq:oglterms}
\end{equation}
where $k+k'=l$ and $s_0 = t$. To get from (\ref{eq:oglterms}) the term $\mathcal{O}(g^{l})$ in Dyson's series we should take the trace of (\ref{eq:oglterms}) and sum it as $\sum_{k=1,\, k'=1; k+k=l}^l$. 

To study (\ref{eq:oglterms}), let us take its leftmost multiplier
\begin{equation}\label{lefthandside}
    \begin{split}
        &\prod_{i=1}^{k}g \int_{0}^{s_{i-1}} ds_i H_I(s_i) = \\
        =&\sum_{\alpha_1 = 1}^{8}...\sum_{\alpha_k = 1}^{8}\prod_{i=1}^{k}g \int_{0}^{s_{i-1}} ds_i h_{\alpha_i}(s_i).
    \end{split}
\end{equation}
Here $\{h_i\}_{i=1}^8$ is the set of all monomials in the interaction Hamiltonian (\ref{full}):
\begin{equation}
\begin{split}
    \{h_i\}_{i=1}^8=\{&a_1a_2^2, a_1^{\dagger}a_2^2,a_1a_2^{\dagger}{}^2,a_1^{\dagger}a_2^{\dagger}{}^2,\\&a_1^2a_2, a_1^{\dagger}{}^2a_2,a_1^2a_2^{\dagger},a_1^{\dagger}{}^2a_2^{\dagger}\}.
\end{split}
\end{equation}
Let us also define the frequency set $\{W_i\}_{i=1}^8$
\begin{equation}
\begin{split}
    \{W_i\}_{i=1}^8 = \{&\omega_1+2\omega_2,-\omega_1+2\omega_2,\omega_1-2\omega_2,-\omega_1-2\omega_2,\\
    &2\omega_1+\omega_2,-2\omega_1+\omega_2,2\omega_1-\omega_2,-2\omega_1-\omega_2\}.
\end{split}
\end{equation}
Keeping in mind the equation $a_{1,2}(s) = a_{1,2} e^{-is\omega_{1,2}}$ let us take one term from the sum (\ref{lefthandside}) corresponding to some $\alpha_1...\alpha_k$:

\begin{equation}\label{eq:termleft}
    \begin{split}
        &\prod_{i=1}^{k}g \int_{0}^{s_{i-1}} ds_i h_{\alpha_i}(s_i) = \prod_{i=1}^{k}g h_{\alpha_i} \times\\&\times \int_{0}^{s_{i-1}} ds_i \exp{\bigg[-is_iW_{\alpha_i}\bigg]}=\\
        &=\prod_{i=1}^{k}g h_{\alpha_i}  \frac{1}{k!}\prod_{i=1}^{k}\int_{0}^{t} ds_i \exp{\bigg[-is_iW_{\alpha_i}\bigg]}.
    \end{split}
\end{equation}
The last step uses the fact that we have $k!$ ways to order $k$ different items and that after taking the operator part out of the integration we get integration of complex valued functions which do not change with ordering. Similarly for the rightmost multiplier of (\ref{eq:oglterms}):
\begin{equation}\label{eq:termright}
    \prod_{i=0}^{k'-1}g h_{\alpha'_{k'-i}} \frac{1}{k'!}\prod_{i=1}^{k'}\int_{0}^{t} ds_i \exp{\bigg[-is_iW_{\alpha'_i}\bigg]}.
\end{equation}

Straightforward calculation shows that the integral terms in (\ref{eq:termleft}) and (\ref{eq:termright}) result in \begin{equation}
    \prod_{i=1}^{k} \frac{(1-e^{-iW_{\alpha_i}t})}{(iW_{\alpha_i})},\;\prod_{i=1}^{k'} \frac{(1-e^{-iW_{\alpha'_i}t})}{(iW_{\alpha'_i})}
\end{equation}

Now we can write (\ref{eq:oglterms}) as
\begin{equation}
    \begin{split}
        &\sum_{\alpha_1=1}^{8}...\sum_{\alpha_k=1}^{8}\sum_{\alpha'_1=1}^{8}...\sum_{\alpha'_{k'}=1}^{8}\frac{g^k}{k!}\frac{g^{k'}}{k'!}\times\\
        &\times\bigg[\prod_{i=1}^{k} h_{\alpha_i} \prod_{i=0}^{k'-1} h_{\alpha'_{k'-i}} \times\\
        &\times\prod_{i=1}^{k} \frac{(1-e^{-iW_{\alpha_i}t})}{(iW_{\alpha_i})}\prod_{i=1}^{k'} \frac{(1-e^{-iW_{\alpha'_i}t})}{(iW_{\alpha'_i})}\bigg].
    \end{split}
\end{equation}

and the equation for $\Delta n_{1,2}$ will be

\begin{equation}
    \begin{split}
        &\Delta n_{1,2}=\sum_{k=0}^{\infty}\sum_{k'=0}^{\infty}\delta_{k+k'}^{0}\sum_{\alpha_1=1}^{8}...\sum_{\alpha_k=1}^{8}\sum_{\alpha'_1=1}^{8}...\sum_{\alpha'_{k'}=1}^{8}\frac{g^k}{k!}\frac{g^{k'}}{k'!}\times\\
        &\times \prod_{i=1}^{k}\frac{(1-e^{-iW_{\alpha_i}t})}{(iW_{\alpha_i})}\prod_{i=1}^{k'} \frac{(1-e^{-iW_{\alpha'_i}t})}{(iW_{\alpha'_i})} \times\\
        &\times {\rm tr} \bigg(\prod_{i=1}^{k} h_{\alpha_i} \rho(0)\prod_{i=0}^{k'-1} h_{\alpha'_{k'-i}}\,\hat{n}_{1,2}\bigg).
    \end{split}
\end{equation}

Here the sum $\sum_{\alpha_1=1}^{8}...\sum_{\alpha_k=1}^{8}\sum_{\alpha'_1=1}^{8}...\sum_{\alpha'_{k'}=1}^{8}$ will have $8^l$ elements for any $l$, so the amount of terms of order $\mathcal{O}(g^l)$ is $8^{l}(l+1)$. This may put doubt in the claim that the higher order $\mathcal{O}(g^l)$ terms of $\Delta n$ can be neglected. However, we believe that it can be done because $8^{l}$ is a huge overestimation; for most $\alpha_1...\alpha_k,\alpha'_1...\alpha'_{k'}$ the trace 
    \begin{equation}\label{eq:termskk'trace}
        {\rm tr} \bigg(\prod_{i=1}^{k} h_{\alpha_i} \rho(0)\prod_{i=0}^{k'-1} h_{\alpha'_{k'-i}}\,\hat{n}_{1,2}\bigg)
    \end{equation}
is zero. Moreover, direct algebraic calculation shows that (\ref{eq:termskk'trace}) is nonzero only if the operator 
    \begin{equation}
        \Theta = \prod_{i=1}^{k} h_{\alpha_i} \prod_{i=0}^{k'-1} h_{\alpha'_{k'-i}}
    \end{equation}
is Hermitian. For example for $l=2$ from 192 terms we get 18 nonzero terms.

\end{document}